\documentclass[graybox]{svmult}

\usepackage{mathptmx}       
\usepackage{helvet}         
\usepackage{courier}        
\usepackage{type1cm}        
\usepackage{makeidx}         
\usepackage{graphicx}        
\usepackage{multicol}        
\usepackage[bottom]{footmisc}

\usepackage{amsmath}
\usepackage{amssymb}
\usepackage{arydshln}
\usepackage{float}
\usepackage{bm}
\usepackage{graphicx}
\usepackage{enumitem} 

\usepackage[numbers]{natbib}
\usepackage{url} 

\begin{document}

\title*{Modelling Career Trajectories of Cricket Players Using Gaussian Processes}
\author{Oliver G. Stevenson and Brendon J. Brewer}
\institute{Oliver G. Stevenson \at University of Auckland, \email{o.stevenson@auckland.ac.nz} \and Brendon J. Brewer \at University of Auckland, \email{bj.brewer@auckland.ac.nz}}
\maketitle

\abstract{
In the sport of cricket, variations in a player's batting ability can usually be measured on one of two scales.
Short-term changes in ability that are observed during a single innings, and long-term changes that are witnessed between matches, over entire playing careers.
To measure long-term variations, we derive a Bayesian parametric model that uses a Gaussian process to measure and predict how the batting abilities of international cricketers fluctuate between innings.
The model is fitted using nested sampling given its high dimensionality and for ease of model comparison.
Generally speaking, the results support an anecdotal description of a typical sporting career.
Young players tend to begin their careers with some raw ability, which improves over time as a result of coaching, experience and other external circumstances.
Eventually, players reach the peak of their career, after which ability tends to decline.
The model provides more accurate quantifications of current and future player batting abilities than traditional cricketing statistics, such as the batting average.
The results allow us to identify which players are improving or deteriorating in terms of batting ability, which has practical implications in terms of player comparison, talent identification and team selection policy.
}

\section{Introduction}

As a sport, cricket is a statistician's dream.
The game is steeped in numerous statistical and record-keeping traditions, with the first known recorded scorecards dating as far back as 1776.
Given the statistical culture that has developed with the growth of cricket, using numeric data to quantify individual players' abilities is not a new concept.
However, despite the abundance of data available, cricket has only recently attracted the attention of statistical analysts in the public realm.
This is potentially due to many previous academic studies being focused on the likes of achieving a fair result ~\citep{duckworthlewis1998, ian2002, carter2004} and match outcome prediction ~\citep{brooks2002, bailey2006, swartz2009, brooker2011}, rather than statistical applications that measure and predict individual player abilities and performances.

For as long as the modern game has existed, a player's batting ability has primarily been recognized using the \textit{batting average}; in general, the higher the batting average, the better the player is at batting.
However, the batting average fails to tell us about variations in a player's batting ability, which can usually be attributed to one of two scales: (1) short-term changes in ability that are observed \textit{during} or \textit{within} a single innings, due to factors such as adapting to the local pitch and weather conditions (commonly referred to as `getting your eye in' within the cricketing community), 
and (2) long-term changes that are observed \textit{between} innings, over entire playing careers, due to the likes of age, experience and fitness levels.

Early studies provided empirical evidence to support the claim that a batsman's score could be modelled using a geometric progression, suggesting players bat with a somewhat constant ability during an innings \citep{elderton1945}.
However, it has since been shown that the geometric assumptions do not hold for many players, due to the inflated number of scores of 0 that are present in many players' career records \citep{kimber1993, bracewell2009}.
Rather than model batting scores, \cite{kimber1993} and \cite{cai2002} used nonparametric and parametric hazard functions respectively, to measure how dismissal probabilities change with a batsman's score.
Estimating a batsman's hazard function, $H(x)$, which represents the probability of getting out on score $x$, allows us to observe how a player's ability varies over the course of an innings.
Both studies found that batsmen appeared to be more likely to get out while on low scores -- early in their innings -- than on higher scores, supporting the idea of `getting your eye in'.

In order to quantify the effects of `getting your eye in', \cite{stevenson2017} proposed an alternative means of measuring how player ability varies during an innings.
The authors use a Bayesian parametric model to estimate the hazard function, allowing for a smooth transition in estimated dismissal probabilities between scores, rather than the sudden, unrealistic jumps seen in \cite{kimber1993} and to a lesser extent \cite{cai2002}.
For the vast majority of past and present international Test players, \cite{stevenson2017} found overwhelming evidence to suggest that players perform with decreased batting abilities early in an innings and improve as they score runs, further supporting the notion of `getting your eye in'.

\subsection{Modelling Between-Innings Changes in Batting Ability}
While there is plenty of evidence to suggest that players do not bat with some constant ability during an innings, it is also unlikely that a player bats with some constant ability throughout their entire career.
Instead, variations in a player's underlying ability are likely to occur \textit{between} innings, due to factors such as how well the player has been performing recently (referred to as `form' in cricket).

If batting form were to have a significant impact on player performance, we should be able to identify extended periods of players' careers with sequences of high scores (indicating the player was `in' form) and sequences of low scores (indicating the player was `out of' form).
On the contrary, \cite{durbach2007} found little empirical evidence to support this idea.
Instead, for the majority of players analyzed in the study, the authors suggest that public perceptions of batting form tend to be overestimated, with many players' scores able to be modelled using a random sequence.

Within a Bayesian framework, \cite{koulis2014} employed the use of a hidden Markov model to determine whether a batsman is in or out of form.
The model estimates a number, $K$, of `underlying batting states' for each player, including the expected number of runs to be scored when in each of the $K$ states.
Parameters that measure: \textit{availability} (the probability a batsman is in form for a given match), \textit{reliability} (the probability a batsman is in form for the next $n$ matches) and \textit{mean time to failure} (the expected number of innings a batsman will play before he is out of form), were also estimated for each batsman.
However, a drawback of this approach is that the model requires an explicit specification of what constitutes an out of form state.
The authors specify a batting state that has a posterior expected median number of runs scored of less than 25, as being out of form, and all other states as being in form.
While in the context of one day or Twenty20 cricket this is not necessarily an unreasonable specification, there are numerous arguments that could be made to justify a low score, scored at a high strike rate, as a successful innings.

In this paper, we extend the Bayesian parametric model detailed in \cite{stevenson2017}, such that we can not only measure and predict how player batting abilities fluctuate during an innings, but also between innings, over the course of entire playing careers.
This allows us to treat batting form as continuous, rather than binary; instead of defining players as `in' or `out' of form, we can describe players as improving or deteriorating in terms of batting ability.
At this stage our focus is on longer form Test and first-class cricket, as limited overs cricket introduces a number of match-specific complications \citep{davis2015}.

\section{Model Specification}

The derivation of the model likelihood follows the method detailed in \cite{stevenson2017}.
If $X \in \{0, 1, 2, 3, ...\}$ is the number of runs a batsman is currently on, we define a \textit{hazard function}, $H(x) \in [0, 1]$, as the probability a batsman gets out on score $x$.
Assuming a functional form for $H(x)$, conditional on some parameters $\theta$, we can calculate the probability distribution for $X$ as follows:

\begin{align} \label{eq:px}
	P(X = x) & = H(x) \prod_{a = 0}^{x - 1} \left[ 1 - H(a) \right].
\end{align}

For any given value of $x$, this can be thought of as the probability of a batsman surviving up until score $x$, then being dismissed.
However, in cricket there are a number of instances where a batsman's innings may end without being dismissed (referred to as a `not out' score).
Therefore, in the case of not out scores, we compute $P(X \geq x)$ as the likelihood, rather than $P(X = x)$.
Comparable to right-censored observations in the context of survival analysis, this assumes that for not out scores the batsman would have gone on to score some unobserved score, conditional on their current score and their assumed hazard function.

Therefore, if $I$ is the total number of innings a player has batted in and $N$ is the number of not out scores, the probability distribution for a set of conditionally independent `out' scores $\{x_i\}_{i = 1}^{I - N}$ and `not out' scores $\{y_i\}_{i = 1}^N$ can be expressed as 

\begin{align} \label{eq:pxy}
	p(\{x\}, \{y\}) & = \prod_{i = 1}^{I - N} \Big(H(x_i) \prod_{a = 0}^{x_i - 1} [1 - H(a)] \Big) \times \prod_{i = 1}^N \Big(\prod_{a = 0}^{y_i - 1} [1 - H(a)] \Big).
\end{align}

When data $\{x, y\}$ are fixed and known, Equation \ref{eq:pxy} gives the likelihood for any proposed form of the hazard function, $H(x; \theta)$.
Therefore, conditional on the set of parameters $\theta$ governing the form of $H(x)$, the log-likelihood is

\begin{align} \label{eq:likelihood}
	\textup{log} \Big(L(\theta) \Big) & = \sum_{i = 1}^{I - N} \textup{log} \ H(x_i) + \sum_{i = 1}^{I - N} \sum_{a = 0}^{x_i - 1} \textup{log} [1 - H(a)] + \sum_{i = 1}^N \sum_{a = 0}^{y_i - 1} \textup{log} [1 - H(a)].
\end{align}

\subsection{Parameterizing the Hazard Function}
The model likelihood in Equation \ref{eq:likelihood} depends on the parameterization of the hazard function, $H(x)$.
As per \cite{stevenson2017}, we parameterize the hazard function in terms of an \textit{effective average function}, $\mu(x)$, which represents a player's ability on score $x$, in terms of a batting average.
Given the prevalence of the batting average in cricket, it is far more intuitive for players and coaches to think of ability in terms of batting averages, rather than dismissal probabilities.
The hazard function can then be expressed in terms of the effective average function, $\mu(x)$, as follows
\begin{equation}
	H(x) = \frac{1}{\mu(x) + 1}
\end{equation}

where the effective average contains three parameters, $\theta = \{\mu_1, \mu_2, L\}$, and takes the following functional form

\begin{equation} \label{eq:mux}
	\mu(x) = \mu_2 + (\mu_1 - \mu_2) \ \textup{exp} \left( \frac{-x}{L} \right).
\end{equation}

Here, $\mu_1$ represents a player's initial batting ability when beginning a new innings, while $\mu_2$ is the player's `eye in' batting ability once used to the specific match conditions.
Both $\mu_1$ and $\mu_2$ are expressed in terms of a batting average.
The timescale parameter $L$, measures the speed of transition between $\mu_1$ and $\mu_2$ and is formally the $e$-folding time.
By definition the $e$-folding time, $L$, signifies the number of runs scored for approximately 63\% (formally $1 - \frac{1}{e}$) of the transition between $\mu_1$ and $\mu_2$ to take place and can be understood by analogy with a `half-life'.
This model specification allows us to answer questions about individual players, such as: (1) how well players perform when they first arrive at the crease, (2) how much better players perform once they have `got their eye in' and (3) how long it takes them to `get their eye in'.

\subsection{Modelling Between-Innings Changes in Batting Ability}

To extend the model further, such that we can measure variations in player batting ability between innings, we use the same likelihood function in Equation \ref{eq:likelihood}. However, we re-parameterize the effective average function to include a time component, $t$, such that
\begin{align}
	\mu(x, t) & = \textup{expected batting average on score } x \textup{, in } t^{th} \textup{ career innings}.
\end{align}

For clarity, we will refer to $\mu(x)$ as the `within-innings' effective average (explaining how ability changes within an innings).
By marginalizing over all scores, $x$, we obtain the `between-innings' effective average, $\nu(t)$, which explains how ability changes between innings, across a playing career.
\begin{align}
	\nu(t) & = \textup{expected batting average in } t^{th} \textup{ career innings}.
\end{align}

When estimating $\nu(t)$, we need to account for variations in ability due to external factors such as: recent form, general improvements/deterioration in skill and the element of randomness associated with cricket.
This is achieved by fitting a $\mu_2$ parameter for each innings in a player's career, where $\mu_{2_t}$ represents a player's `eye in' batting ability, corresponding to their $t^{th}$ innings.
We are then able to predict the expected batting average in each innings, $\nu(t)$, analytically using Equation \ref{eq:mux}.

To afford a player's underlying batting ability a reasonable amount of flexibility, the set of $\{\mu_{2_t} \}$ terms are modelled using a Gaussian process.
A Gaussian process is fully specified by an underlying mean value, $m$, and covariance function, $K(X_i, X_j)$, which will determine by how much a player's batting ability can vary from innings to innings \citep{rasmussenwilliams2006}.
Our choice of covariance function is the commonly used squared exponential covariance, which contains scale and length parameters $\sigma$ and $\ell$.

Therefore, the model contains the set of parameters $\theta = \{\mu_1, \{\mu_{2_t}\}, L, m, \sigma, \ell \}$.
The model structure with respect to parameters $\mu_1$, $L$, $C$ and $D$ follows the model specification detailed in \cite{stevenson2017}, with the parameters assigned the following prior distributions.
\begin{equation*} 
	\begin{aligned}[c]
  		\hspace{5em} \mu_1 \leftarrow C\mu_2 \\
 		L \leftarrow D\mu_2 \\
  		C \sim \textup{Beta}(1, 2) \\
  		D \sim \textup{Beta}(1, 5) \\
	\end{aligned}
	\begin{aligned}[c]
  		\hspace{5em} \textup{log}(\mu_{2_t}) & \sim \textup{Gaussian process}(m, K(X_i, X_j; \sigma, \ell)) \\
		m & \sim \textup{Lognormal}(\textup{log}(25), 0.75^2) \\
		\sigma & \sim \textup{Exponential}(\textup{10}) \\
		\ell & \sim \textup{Uniform}(0, 100)
	\end{aligned}
\end{equation*}

These priors are either non-informative or are relatively conservative, while loosely reflecting our cricketing knowledge.
It is worth noting, that as we are measuring ability in terms of a batting average (which must be positive), we model log($\mu_{2_t}$), rather than just $\mu_{2_t}$, to ensure positivity in our estimates.

As the model requires a set of  $\{\mu_{2_t}\}$ parameters to be fitted (one for each innings played), the model can contain a large number of parameters for players who have enjoyed long international careers.
Therefore, to fit the model we employ a C++ implementation of the nested sampling algorithm \citep{skilling2006}, which uses Metropolis-Hastings updates and is able to handle both high dimensional and multimodal problems that may arise.
The model output provides us with the posterior distribution for each of the model parameters, as well as the marginal likelihood, which makes for trivial model comparison.
For each player analyzed, we initialize the algorithm with 1000 particles and use 1000 MCMC steps per nested sampling iteration.

\section{Analysis of Individual Players}
\subsection{Data}

The data we use to fit the model are simply the Test career scores of an individual batsman
and are obtained from Statsguru, the cricket statistics database on the Cricinfo website\footnote{www.espncricinfo.com}.
As the model assumes that a player's ability is not influenced by the specific match scenario, it is best suited to longer form cricket, such as Test matches, where there is generally minimal external pressure on batsmen to score runs at a prescribed rate.

\subsection{Modelling Between-Innings Changes in Batting Ability}
\begin{figure}[h]
	\centering
	\includegraphics[width = 0.75\linewidth]{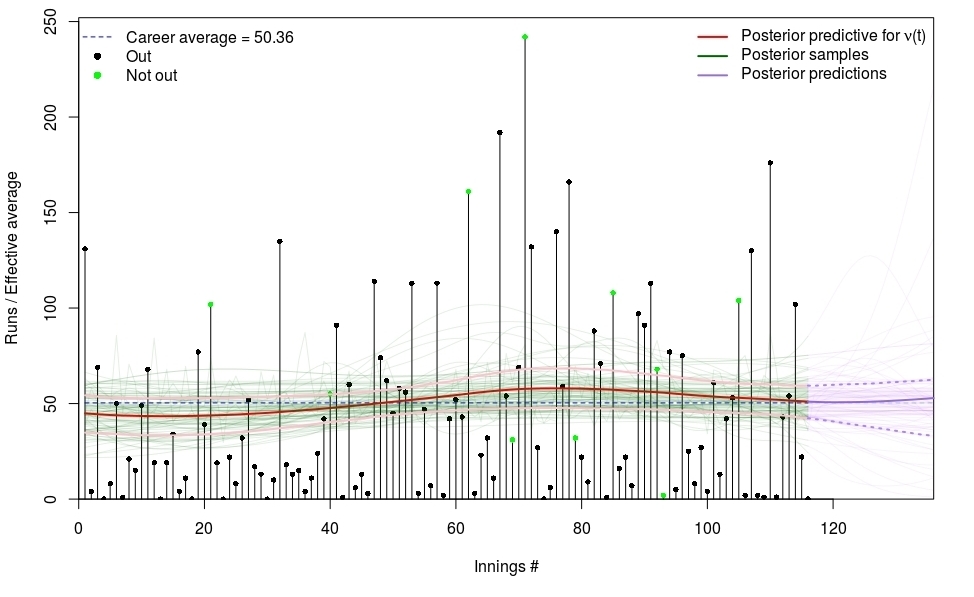}
	\caption{Posterior predictive effective average function for $\nu(t)$ (red), fitted to Kane Williamson's Test career data, including a subset of posterior samples (green), future predictions (purple) and a 68\% credible interval (pink/dotted purple).}
	\label{figure:GPWilliamson}
\end{figure}

To illustrate the practical implications of the model, let us consider the Test match batting career of current New Zealand captain, Kane Williamson.
The evolution of Williamson's between-innings effective average, $\nu(t)$, is shown in Figure \ref{figure:GPWilliamson} and suggests that early in his career, Williamson was not as good a batsman as he is today.
In fact, it was not until playing in roughly 50 innings that he began to consistently bat \textit{at least} as well as his current career average of 50.36.
This is not surprising, as it is a commonly held belief that many players need to play in a number of matches to `find their feet' at the international level, before reaching their peak ability.

\begin{figure}[h]
	\centering
	\includegraphics[width = 0.75\linewidth]{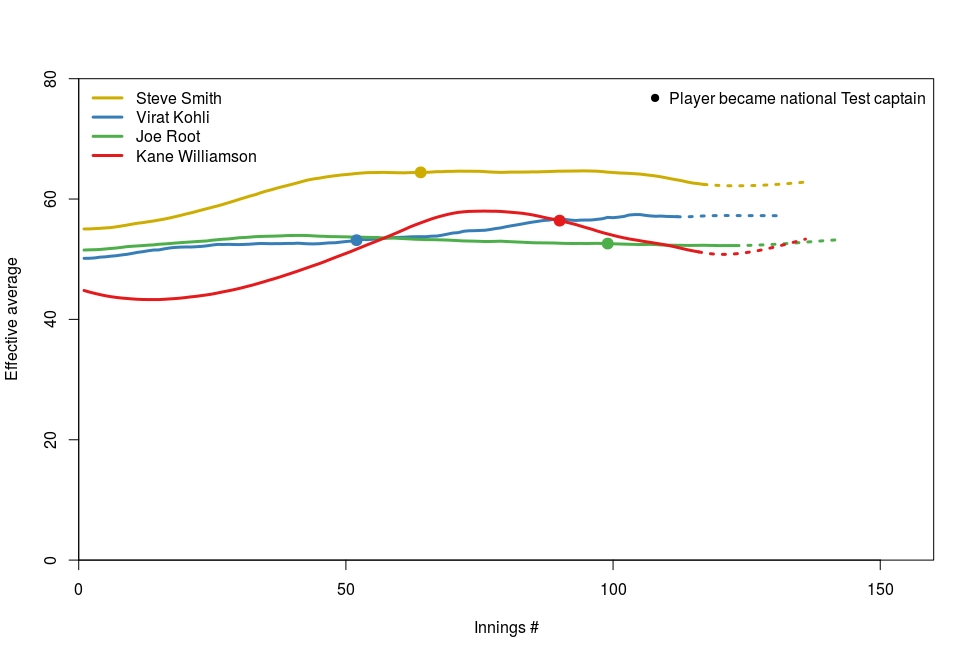}
	\caption{Posterior predictive effective average functions, $\nu(t)$, for `the big four', including predictions for the next 20 innings (dotted).}
	\label{figure:GPAll}
\end{figure}

To gain a better understanding of how Williamson compares to other batsmen globally, we can compare multiple players' effective average functions.
Figure \ref{figure:GPAll} compares the predictive effective average functions for the current top four batsmen in the world, as ranked by the official International Cricket Council (ICC) ratings\footnote{As of 1$^{st}$ August, 2018: (1) Steve Smith, (2) Virat Kohli, (3) Joe Root and (4) Kane Williamson -- commonly referred to as 
`the big four'}.

\vspace*{-\baselineskip}

As we might expect, all players appear to have improved in terms of batting ability since the start of their careers.
Again, this supports the concept of `finding your feet' at the international level, although different players appear to take different lengths of time to adjust to the demands of international cricket.

Table \ref{table:compare} shows each player's predicted effective average for their next innings, as well as their ICC rating.
The order of these four players remains unchanged when ranking by predicted effective averages instead of ICC ratings, however, as we have computed the posterior predictive distributions for $\nu(t)$, our model has the added advantage of being able to quantify the differences in abilities between players.
Rather than concluding `Steve Smith is 26 rating points higher than Virat Kohli', we can make more useful statements such as: `we expect Steve Smith to outscore Virat Kohli by 5.1 runs in their next respective innings' and `Steve Smith has a 68.8\% chance of outscoring Virat Kohli in their next respective innings'.

\vspace*{-\baselineskip}
\begin{table}[h]
\caption{Predicted effective averages, $\nu(t)$, for the next career innings for `the big four'. The official ICC Test batting ratings (as of 1$^{st}$ August, 2018) are shown for comparison.}
\centering
\begin{tabular}{l c c r}
\hline
\textbf{Player} & \textbf{Career average} & \textbf{Predicted} $\mathbf{\nu}$\textbf{(next innings)} & \textbf{ICC Rating (\#)} \\
\hline
S. Smith (AUS) & 61.4 & 62.5 & 929 (1) \\		
V. Kohli (IND) & 53.4 & 57.4 & 903 (2) \\
J. Root (ENG) & 52.6 & 52.6 & 855 (3) \\
K. Williamson (NZ) & \hspace{3em}50.4\hspace{3em} & \hspace{7em}51.2\hspace{7em} & 847 (4) \\
\hline
\end{tabular}
\label{table:compare}
\end{table}
\vspace*{-\baselineskip}
\vspace*{-\baselineskip}

\section{Concluding Remarks and Future Work}

We have presented a novel and more accurate method of quantifying player batting ability than traditional cricketing statistics, such as the batting average.
The results provide support for the common cricketing belief of `finding your feet', particularly for players beginning their international careers at a young age, with many batsmen taking a number of innings to reach their peak ability in the Test match arena.
With respect to batting form, the model appears to reject the idea of recent performances as having a significant impact on innings in the near future.
In particular, it appears that the effect of recent form varies greatly from player to player.

A major advantage of the model is that we are able to maintain an intuitive cricketing interpretation, allowing for the results and implications to be easily digested by coaches and selectors, who may have minimal statistical training.
Additionally, we are able to make probabilistic statements and comparisons between players, allowing us to easily quantify differences in abilities and predict the real life impacts of selecting one player over another.
As such, the findings have practical implications in terms of player comparison, talent identification, and team selection policy.

It is worth noting that we have ignored important variables, such as the number of balls faced in each innings, as well as the strength of the opposition.
Currently, the model treats all runs scored equally.
Implementing a means of incorporating more in-depth, ball-by-ball data and including the strength of opposition bowlers will reward players who consistently score highly against world-class bowling attacks.

\bibliographystyle{spmpsci}
\bibliography{references}

\end{document}